\documentclass[times,10pt,twocolumn]{article}

\usepackage{latex8} 
\usepackage{times} 
\usepackage{epsfig} 
\usepackage{amssymb}  
\usepackage{latexsym}  
\usepackage[PostScript=dvips]{diagrams}
\newarrow{Toto}<--->
\newarrow{Is}=====
\newarrow{Dotsto}....>
\newarrow{Dots}.....

\begin{document}  

\newcommand{\bpf}{\noindent {\bf Proof:} }
\def\endproof{\hfill$\Box$}

\newcommand{\cnot}{\textsc{\footnotesize CNOT}}
\newcommand{\bit}{\begin{itemize}}
\newcommand{\eit}{\end{itemize}\par\noindent}
\newcommand{\ben}{\begin{enumerate}}
\newcommand{\een}{\end{enumerate}\par\noindent}
\newcommand{\beq}{\begin{equation}}
\newcommand{\eeq}{\end{equation}\par\noindent}
\newcommand{\beqa}{\begin{eqnarray*}}
\newcommand{\eeqa}{\end{eqnarray*}\par\noindent}
\newcommand{\beqn}{\begin{eqnarray}}
\newcommand{\eeqn}{\end{eqnarray}\par\noindent}
\newcommand{\suck}{\vspace{-2.25mm}}

\newcommand{\dd}{\llcorner}
\newcommand{\sdot}{\bullet}
\newcommand{\ddd}{\lrcorner}
\newcommand{\uu}{\ulcorner}
\newcommand{\uuu}{\urcorner}
\newcommand{\ot}{\otimes}

\newcommand{\HH}{\mathcal{H}}
\newcommand{\KK}{\mathcal{K}}
\newcommand{\ie}{\textit{i.e.}\ }
\newcommand{\Zero}{\mathbf{0}}
\newcommand{\CC}{\mathbf{C}}
\newcommand{\II}{{\rm I}}
\newcommand{\PP}{{\rm P}}
\newcommand{\bra}{\mathsf{bra}}
\newcommand{\FdHilb}{\mathbf{FdHilb}}

\newcommand{\neoiota}{\psi}

\newtheorem{Th}{Theorem}[section]
\newtheorem{theorem}[Th]{Theorem}
\newtheorem{proposition}[Th]{Proposition}
\newtheorem{lemma}[Th]{Lemma}
\newtheorem{corollary}[Th]{Corollary}
\newtheorem{definition}[Th]{Definition}
\newtheorem{example}[Th]{Example}
\newcommand{\measure}[2]{\mathsf{measure} \; #1 \; \mathsf{in} \; #2}
\newcommand{\labarrow}[1]{\stackrel{#1}{\longrightarrow}}
\newcommand{\Skip}{\mathsf{skip}}
\newcommand{\DS}[1]{\mathcal{D}\lsem #1 \rsem}
\newcommand{\WP}[1]{\mathrm{WP}\lsem #1 \rsem}
\newcommand{\DP}[1]{\mathsf{DProb}(#1)}
\newcommand{\Prob}[1]{\mathbf{P}(#1)}
\newcommand{\pfn}{\rightharpoonup}
\newcommand{\St}{\mathcal{S}}
\newcommand{\lsem}{\llbracket}
\newcommand{\rsem}{\rrbracket}

\title{High-Level Methods for Quantum Computation and Information}

\author{Samson Abramsky\\ 
Oxford University Computing Laboratory}

\maketitle 
\thispagestyle{empty}

\Section{Background}

Quantum information and computation is concerned with the use of
quantum-mechanical systems to carry out computational and
information-processing tasks \cite{Nielsen}.
In the few short years that this approach has been studied, a number of
remarkable concepts and results have emerged, most notably:
\begin{itemize}
\item A couple of spectacular algorithms --- Shor's
polynomial-time algorithm for prime factorization \cite{Shor2} and
Grover's
sub-linear search algorithm \cite{Grover}.
\item A number of information protocols, exemplified by \emph{quantum
teleportation}, which exploit quantum entanglement in an essential
fashion. We give a thumbnail sketch of teleportation here, since it may
be less familiar, and it will serve as a useful motivating example.
Teleportation uses just two classical bits to transport an unknown
qubit $q$ from one site to another. Since a qubit is specified by an
arbitrary pair of complex
numbers $(\alpha ,\beta)$ satisfying $|\alpha|^2 + |\beta|^2 = 1$,
achieving this information transfer with just two
classical bits is no mean feat! It is accomplished by using an
entangled pair $q_A$, $q_B$ of qubits, one held at the source site $A$
and one at the
target $B$, as a `quantum information channel', and using a measurement
performed on $q$ and $q_A$ at $A$ to
cause a `collapse' in $q_B$ at $B$. The two classical bits are used to
tell
the target site $B$ what the outcome of the measurement performed at $A$
was; a
`correction' operation can then be performed at $B$, after
which the state of $q_B$ will be equal to the original state
of $q$.
(Because of the measurement, the input qubit no longer
has this state --- the information in the source has been `destroyed' in
transferring it to the target).
\end{itemize}

Teleportation is simply the most basic of a family of quantum
protocols, including \emph{logic-gate teleportation}
\cite{Gottesman}, \emph{entanglement swapping} \cite{Swap}, and
\emph{quantum key exchange} \cite{Ekert}, which form the basis for novel
and potentially very
important applications to secure and fault-tolerant communication and
computation \cite{BEZ,Gottesman,Nielsen,Preskill,Shor}.

\subsection*{The need for high-level methods}

The current tools available for developing quantum algorithms and
protocols are deficient on two main levels.
\begin{itemize}
\item Firstly, they are too \emph{low-level}. Quantum algorithms are
currently mainly described using the `network model' corresponding
to circuits in classical computation. One finds a plethora of ad hoc
calculations with `bras' and `kets', normalizing constants, matrices
etc. The arguments for the benefits of a high-level, conceptual
approach to designing, programming and reasoning about quantum
computational systems are just as compelling as for classical
computation. Moreover, there is the whole issue of integrating
quantum and classical features, which would surely be mandatory in
any practicable
system.

\item At a more fundamental level, the standard mathematical framework
for quantum mechanics (which is
essentially due to von Neumann \cite{vN}) is actually
\emph{insufficiently comprehensive} for informatic purposes.
In describing a protocol such as teleportation, or any quantum
process in which \emph{the outcome of a measurement is used to determine
subsequent actions}, the von Neumann formalism does not capture the flow of
information from the classical or macroscopic level, where the results 
of measurements of the quantum-mechanical system are recorded, back to the
quantum level. This flow, and the accompanying use of `classical
information', which plays a key role in protocols such as
teleportation,  must therefore be handled informally.
As quantum protocols and computations grow
more
elaborate and complex, this point is likely to prove of increasing
importance.
\end{itemize}

\Section{Recent Progress}

In joint work with Bob Coecke, reported in this Conference Proceedings 
\cite{AbrCoe2}, we have recently made some striking progress in addressing both these
points.
We have recast the von Neumann formalism at a more abstract and
conceptual level, using category theory. This enables a high-level
but effective approach to modelling and reasoning about the key
features of quantum information processing, including preparation
and measurement of entangled states, unitary operations and
classical communication. The effectiveness of these methods is shown
by the detailed treatment given in \cite{AbrCoe2} of three of the main
quantum protocols: teleportation, logic-gate teleportation (which is
universal for quantum computation), and entanglement
swapping. Because of the explicit treatment of `classical
communication' --- \ie the use of measurement outcomes to determine
subsequent actions, possibly elsewhere in a compound system than the
site at which the measurement was performed --- it can reasonably be
claimed that these are the \emph{first} completely formal
descriptions and proofs of correctness of these protocols. Moreover,
the correctness proofs are themselves at a high level, using (and
re-using) key
structural lemmas which are valid in wide generality.

One of the main results in \cite{AbrCoe2} is a complete formal
description---including the classical communication--- of the
teleportation protocol within a purely categorical
semantics. This
semantics is sufficiently strong to prove correctness of the
protocol.

Consider the following diagram:
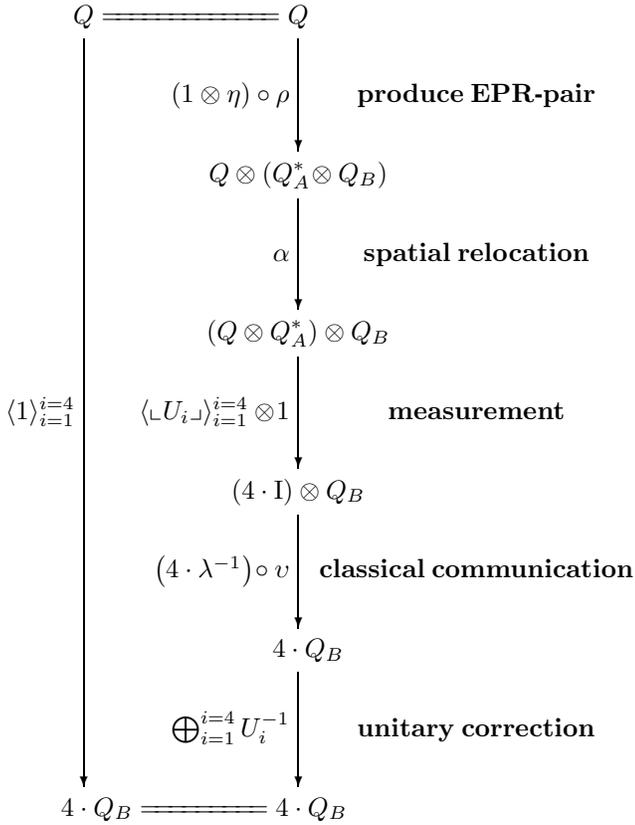
\begin{diagram}
Q&\rIs&Q\\
&&\dTo^{(1\otimes \eta)\circ\rho}&\hspace{-1.5cm}{\bf produce\
EPR\mbox{\bf
-}pair}\\ &&Q\otimes(Q_A^*\!\otimes Q_B)\\
&&\dTo^{\alpha}&\hspace{-1.5cm}{\bf spatial\ relocation}\\
&&(Q\otimes Q^*_A)\otimes Q_B\\
\dTo^{\langle 1\rangle_{i=1}^{i=4}}&&\dTo^{\quad\quad\langle \dd
U_i\ddd\rangle_{i=1}^{i=4}\otimes\! 1}&\hspace{-1.5cm}{\bf
measurement}\\ &&\left(4\cdot{\rm
I}\right)\otimes Q_B\\
&&\dTo^{\left(4\cdot\lambda^{-1}\right)\!\circ\upsilon}&\hspace{-1.5cm}{\bf
\quad classical\ communication\quad}\\ &&\ \ \ 4\cdot Q_B\\
&&\dTo^{\bigoplus_{i=1}^{i=4}U_i^{-1}}&\hspace{-1.5cm}{\bf
unitary\ correction}\\
\ \ \ \ 4\cdot Q_B&\rIs&4\cdot Q_B\!\!\!\!\!\!
\end{diagram}
Here $Q$, $Q_A$ and $Q_B$ are all instances of a qubit
  object, $\mathrm{I}$ is the tensor unit, and $n \cdot A$ is the
  biproduct of $n$ copies of $A$. The natural isomorphisms are
$\rho:Q\simeq Q\otimes\II$, $\lambda:Q\simeq\II\otimes Q$, `associativity'
$\alpha$ and
`distributivity' $\upsilon$ of tensor $\otimes$ over biproduct $\oplus$.
The morphism
$\eta:\II\to Q^*\otimes Q$ is the unit of compact closure, that is the
`name' of $1_Q:Q\to Q$,
and represents preparation of the EPR state \cite{Nielsen}, while
$\dd U_i\ddd: Q\otimes Q^*\to\II$ is the `coname' of $U_i$, and the
tupling $\langle \dd
U_i\ddd\rangle_{i=1}^{i=4}$ represents a Bell-base measurement
\cite{Nielsen}.

The right-hand-side of the diagram gives a complete description of the teleportation protocol, while the
left-hand-side expresses the intended behaviour (copying the qubit
from $Q$ to $Q_B$).  The proof of
correctness, that is the commutativity of the diagram, can be found in
\cite{AbrCoe2}.

The abstract setting is that of \em strongly compact closed
categories with biproducts\em. Any such category allows us define abstract
counterparts to the
basic ingredients of quantum mechanics such as measurement and unitary
data transformation, and to
add to that a description of classical communication.

Conceptually:
\begin{itemize}
\item the tensor product
$\otimes$ of the monoidal structure
allows compound systems to be described;
\item the compact closed structure
(cf.~`names' and `conames')
allows preparations and
measurements of entangled states to be represented, \emph{and their
key properties to be proved};
\item the biproducts allows measurements, branching on measurement
  outcomes, superpositions, and classical
communication (using distributivity of tensor over biproduct) to be
captured.
\end{itemize}
Although these axioms are all structural, and seem purely
`qualitative', in fact they suffice to yield good notions of
`scalars', `probability amplitudes', and the \emph{Born rule}---the key
quantitative feature of quantum mechanics.

\paragraph{Entanglement as information flow.} A key part of our work
is the
analysis of the information flow inherent in entanglement, which
exploits the compact closed structure.
In the above commutative diagram for teleportation, note that the
measurement is formed by tupling the \emph{conames} $\dd U_i\ddd$ of (the
inverses
of) the unitary correction operators $U_i^{-1}\!$. This makes visible the
structure underlying the apparently ad hoc juggling with Bell bases and
unitary matrices in the standard presentations. The flow along each
`branch' as we follow the possible measurement outcomes can then be
analyzed
using general algebraic properties of compact closed categories. See
\cite{AbrCoe2} for details, and also \cite{Coe1,Coe2} for an extended
account at a more concrete level, with
many diagrams and examples.

\Section{Some Further Developments}

We survey some promising further developments we are currently
pursuing.

\paragraph{Categorical quantum logic.}
Although the work in \cite{AbrCoe2} is not presented in logical terms, the
well-established paradigm of categorical logic and proof theory is
directly applicable, and leads to another perspective, and some
potentially very useful syntactic methods.
In particular, my student Ross Duncan and I are currently studying a
notion of \emph{proof nets} for compact closed categories with
biproducts. This builds on
the
seminal study by Kelly and Laplaza of
coherence for compact closed categories \cite{KellyLaplaza}, to yield an
amenable
graphical proof theory corresponding to the categorical semantics in
\cite{AbrCoe2}.
This offers the prospect that the correctness proofs in \cite{AbrCoe2} can
be
performed automatically by \emph{cut-elimination}. Duncan is building
an experimental implementation of proof nets, and we intend to use it to
perform some of
these
computations.

Since the quantitative features of quantum mechanics (scalars, the
Born rule etc.) are also captured in the categorical semantics in
\cite{AbrCoe2}, it seems that one can obtain an interesting  and applicable
diagrammatic tool for automating a class of `structural calculations'
in quantum informatics.

\paragraph{Transformations for parallelism and fault-tolerance.}
In \cite{Coe1,Coe2}, Coecke uses the Logic of Entanglement setting, which
is
now subsumed by the more general categorical semantics in \cite{AbrCoe2},
to
give a `compilation scheme' for quantum networks into a special,
highly parallel form. (A similar compilation is carried out by Duncan
in \cite{Duncan}, using the proof-net formalism.)
This parallel form, which avoids reuse of outputs from quantum gates,
is particularly attractive because of its \emph{fault-tolerance
properties}; these have been emphasized in the influential work
of Shor \cite{Shor} and Preskill \cite{Preskill}.
We aim to pursue these ideas, and more generally to see how our powerful
algebraic methods can be applied to yield useful program transformations
for quantum circuits and other computational systems. The situation is
quite analogous to that in standard program transformation and
hardware design and verification, where categorical methods have been
applied with considerable success
\cite{HardwDesignVerif1,HardwDesignVerif2}.

\paragraph{The One-Way Quantum Computation Model.}
The `one-way' or `measurement-based' model \cite{Briegel,Brie} has been
proposed
recently as an alternative to the standard `network model' of quantum
circuits.
In the one-way model, the computation starts from an initially
prepared entangled state (a `cluster state'), which consists of
entangled qubits laid out on
a 2-D grid, or more generally a graph. Computation proceeds by
performing measurements on these qubits. In general, these have to be
combined with unitary corrections. Thus the arrangement generalizes
that of teleportation. This model is seen as promising because of its
good properties as regards modularity and fault-tolerance.

The full expressive power of the model has yet to be explored; it may
well offer new possibilities going beyond the standard network model
(which it has been shown can be represented within it).

The categorical semantics in \cite{AbrCoe2} seems well adapted to study
the
one-way model. In fact, the key modularity property of the one-way
model appears to fall out as a consequence of one of our general
algebraic results --- which in turn corresponds to the soundness of
Cut-elimination in the categorical logic approach described
above. Our aim here will be to explore how our methods can be used
to give an analytical description of the one-way model, and how this
relates to current methods based on the stabiliser formalism
\cite{Nielsen}. We
believe that our algebraic approach will lead to simpler and more tractable
descriptions of this model, and more insight into its expressive
power.

\paragraph{Multipartite entanglement.}
The methods and results in \cite{AbrCoe2} give a
comprehensive semantic and logical analysis of bipartite
entanglement. The situation with multipartite entanglement ---
several qubits or other quantum systems mutually entangled --- is less
clear. It seems that some of the relevant structure will be
addressed by our categorical logic and semantics, since we can
represent compound quantum systems of any degree, arbitrary
(linear) functional dependencies, etc. Whether this suffices to
address all significant forms of multipartite entanglement remains to
be investigated. This is an area which is generally agreed to be
both important and very poorly understood currently, so if logical methods
do gain some traction this would have considerable impact.

\paragraph{Foundational issues.}
There are numerous important and promising ideas which arise in
seeking to extend and strengthen the categorical foundations
established in \cite{AbrCoe2}. Firstly, the discussion in \cite{AbrCoe2}
is limited
to \emph{Finitary Quantum Mechanics}, in which only
finite-dimensional `spaces' (concrete or abstract), corresponding to
observables with finite spectra, are considered.
A first step towards the general case has already been taken in the
previous work by Abramsky, Blute and Panangaden on Nuclear ideals
\cite{ABP}, which shows how to lift the compact closed arguments to
the general case. However, a proper
treatment of observables with continuous
spectra is challenging.
We have started collaborative work with Rick Blute and
Prakash Panangaden on this topic.

Another important issue is to see how far our general and axiomatic
approach can be exploited to yield insights into the degrees of
freedom in quantum mechanics, and to what extent the structure is
\emph{forced} by various information-theoretic principles.
We have already shown in \cite{AbrCoe2} how a kind of `reverse
mathematics'
applies, in which one can show what requirements are placed on the
ambient category in order for protocols such as teleportation
to be expressed.
We will also
look
at possible weakenings of the axioms, and how much quantum
information processing can still be carried out in weaker settings.

\bibliographystyle{latex8}  

\begin{thebibliography}{99}
\bibitem{ABP}
S.~Abramsky, R.~Blute, and P.~Panangaden. Nuclear and trace ideals in
tensored $*$-categories. {\it Journal of Pure and Applied Algebra} {\bf
143},
3--47 (1999).\suck
\bibitem{AbrCoe1}  S.~Abramsky and B.~Coecke.  Physical traces:
classical vs.~quantum information  processing. {\it
Electronic notes on Theoretical Computer  Science} {\bf 69} (2003) --
CTCS`02 issue. \texttt{arXiv:cs/0207057}\suck
\bibitem{AbrCoe2}
S.~Abramsky and B.~Coecke.
A categorical semantics of quantum protocols.
{\it Proceedings of 19th Annual IEEE Symposium on Logic in
Computer Science} (LiCS`04).
\texttt{arXiv:quant-ph/0402130}\footnote{Papers posted on the
physics
\texttt{arXiv}'s
are downloadable at the address \texttt{www.arXiv.org/{\it name}}
e.g.~\texttt{www.arXiv.org/quant-ph/0402130}.}\suck
\bibitem{InteractionCats}
S.~Abramsky, S.~J.~Gay and R.~Nagarajan.
Interaction categories and foundations of typed concurrent programming.
{\em Deductive Program Design: Proceedings of the 1994 Marktoberdorf
International
Summer School}, pp.~35--113. NATO ASI Series F, Springer-Verlag,
1995.\suck
\bibitem{Bae}
J.~Baez. Quantum Quandaries: A Category-Theoretic Perspective.
To appear in \textit{Structural Foundations of Quantum Gravity},
ed. S. French, D. Rickles and J. Sahatsi, Oxford University Press.
Available at \texttt{http://math.ucr.edu/home/baez/quantum/}.\suck
\bibitem{BBC}
C.~H.~Bennett, C.~Brassard, C.~Cr\'epeau, R.~Jozsa, A.~Peres and
W.~K.~Wooters.  Teleporting an unknown quantum state via dual classical
and Einstein-Podolsky-Rosen channels.  {\it Physical Review Letters} {\bf
70}, 1895--1899 (1993).\suck
\bibitem{HardwDesignVerif1}
R.~S.~Bird and O.~de Moor.
{\it Algebra of Programming}. Prentice-Hall International, 1996.\suck
\bibitem{BEZ}
D.~Bouwmeester, A.~Ekert and A.~Zeilinger.
\em The Physics of Quantum Information\em .
Springer-Verlag, 2001.\suck
\bibitem{Coe1}
B.~Coecke.
The Logic of entanglement. An invitation. PRG-RR-03-12 Oxford University
Computing Laboratory.
\texttt{http://web.comlab.ox.ac.uk/oucl/
publications/tr/rr-03-12.html}.\suck
\bibitem{Coe2}
B.~Coecke. The logic of entanglement. \texttt{arXiv:quant-
ph/0402014}\suck
\bibitem{Duncan}
R.~Duncan.  D.Phil. thesis transfer report. Oxford University
Computing Laboratory, 2004.\suck
\bibitem{Ekert}
A. K. Ekert,
Quantum cryptography based on Bell's theorem.
{\em Physical Review Letters} {\bf 67}, 661--663 (1991).\suck
\bibitem{Gottesman}
D.~Gottesman and I.~L.~Chuang.
Quantum teleportation is a universal computational primitive.  {\it
Nature} {\bf 402}, 390--393 (1999). \texttt{arXiv:quant-ph/9908010}\suck
\bibitem{Grover}
L.~K.~Grover.
Quantum mechanics helps in searching for a needle in a haystack.
{\em Physical Review Letters} {\bf 79}, 325--328 (1997).\suck
\bibitem{HardwDesignVerif2}
G.~Jones and M.~Sheeran.
Circuit design in Ruby.
{\it Formal methods for VLSI design}, ed. J. Staunstrup, 13--70, Elsevier,
1993.\suck
\bibitem{KellyLaplaza}
G.~M.~Kelly. and M.~L.~Laplaza. Coherence for compact closed categories.
{\it Journal of Pure and Applied Algebra} {\bf 19}, 193--213 (1980).\suck
\bibitem{vN}
J.~von Neumann. {\it Mathematische Grundlagen der
Quantenmechanik}. Springer--Verlag, 1932. English translation:
{\it Mathematical Foundations of Quantum Mechanics}.
Princeton University Press, 1955.\suck
\bibitem{Nielsen}
M.~A.~Nielsen and L.~Chuang.
\em Quantum Computation and Quantum Information\em.
Cambridge University Press, 2000.\suck
\bibitem{Preskill}
J.~Preskill.
Reliable quantum Computers. \texttt{arXiv: quant-ph/9705031}\suck
\bibitem{Briegel}
R.~Raussendorf and H.~J.~Briegel.
Computational model for the one-way quantum computer: concepts and
summary. \texttt{arXiv:quant-ph/0207183}\suck
\bibitem{Brie}
R.~Raussendorf, D.~E.~Browne and H.~J.~Briegel. Measurement-based quantum
computation
on cluster states. \texttt{arXiv:quant-ph/0301052}\suck
\bibitem{Selinger}
P.~Selinger. Towards a quantum programming language.
{\em Mathematical Structures in Computer Science}. To appear.\suck
\bibitem{Shor2}
P.~W.~Shor.
Algorithms for quantum computation:
Discrete logarithms and factoring.
{\em Proceedings of the 35th Annual Symposium on the Foundations of
Computer
Science}, pp.~124--134. IEEE Computer Society Press, 1994.\suck
\bibitem{Shor}
P.~W.~Shor.
Fault-tolerant quantum computation. {\it Proceedings of the 37nd Annual
Symposium on Foundations of Computer Science}, pp.~56--65. IEEE Computer
Society Press, 1996. \texttt{arXiv:quant-ph/9605011}\suck
\bibitem{vanTonder}
A.~van Tonder.
Quantum computation, categorical semantics,
and linear logic. \texttt{arXiv:quant-ph/0312174}\suck
\bibitem{WZ}
W.~Wootters and W.~Zurek. A single
quantum cannot be cloned. {\it Nature} {\bf 299}, 802--803 (1982).\suck
\bibitem{Swap}
M.~\.{Z}ukowski, A.~Zeilinger, M.~A.~Horne and A.~K.~Ekert.
`Event-ready-detectors' Bell experiment via entanglement swapping. {\it
Physical Review Letters} {\bf 71}, 4287--4290 (1993).\suck
\end{thebibliography}

\end{document}